\documentstyle{article}
\date{}
\title{Existence of Hyperbolic Orbits for N-body Type Problems\footnote{Supported partially by NSF of China.}}
\author{{Donglun Wu\footnote{Corresponding author. {\em E-mail address:} wudl2008@163.com(D. Wu)}, Shiqing Zhang}\\
{\small Department of Mathematics, Sichuan University,}\\
{\small Chengdu 610064, People's Republic of China}}
\begin{document}
\maketitle \large \baselineskip 14pt

\begin{quote}

{\bf Abstract}\ \ In this paper, we use variational minimizing
method to prove the existence of hyperbolic solution with a
prescribed positive energy for N-body type problems with strong
forces. Firstly, we get periodic solutions using suitable
constraints, then by taking limit about a sequence of periodic
solutions, we get the hyperbolic orbits.



\end{quote}

\section{Introduction and Main Results}

\ \ \ \ \ \ In this paper, we consider the following N-body problems
\begin{equation}
   m_{i}\ddot{u}_{i}(t)+\nabla_{u_{i}} V(u_{1}(t),\cdots,u_{N}(t))=0,\ \ \ (1\leq i\leq N),\label{1}
\end{equation}
with
\begin{equation}
   \frac{1}{2}\sum_{i=1}^{N} m_{i}|\dot{u}_{i}(t)|^{2}+ V(u_{1}(t),\cdots,u_{N}(t))=H\label{2}.
\end{equation}

In 1686, Newton found the universal gravity law. In his classical
book, {\em Mathematical methods of natural philosophy}, he solved
the elliptical orbits for 2-body problem. Referring to the two-body
problem which can be reduced to center force problem with a center
potential $V(x)=-\displaystyle\frac{1}{|x|}$, it is well known that
\begin{eqnarray*}
&&(\mbox{i}).\ \mbox{If H$<$0, the solution for systems (\ref{1}) and (\ref{2}) is an elliptic orbit;}\nonumber\\
&&(\mbox{ii}).\ \mbox{If H=0, the solution for systems (\ref{1}) and (\ref{2}) is a parabolic orbit;}\nonumber\\
&&(\mbox{iii}).\ \mbox{If H$>$0, the solution for systems (\ref{1})
and (\ref{2}) is a hyperbolic orbit.}
\end{eqnarray*}

Using variational methods, many mathematicians tried to prove the
existence of periodic orbits and unbounded orbits for N-body-type
Hamiltonian systems(\cite{1,18,20,23,12,17,2,3,7,9,10,25,11,16} and
the references therein). Using the Mountain Pass Lemma, Ambrosetti
and Coti Zelati \cite{18} studied the existence of weak solutions
for symmetrical N-body problems with any given masses
$m_{1},\cdots,m_{N}>0$ and fixed energy $H<0$. Recently, E. Maderna
and A. Venturelli proved the existence of global parabolic orbits
for Newtonian N-body problem. They proved the following theorem.

\vspace{0.3cm}{\bf Theorem A(See\cite{7})}\ \  {\em Given any
initial configuration $y_{i}$ and any minimizing normalized central
configuration $y_{0}$, there exists a parabolic solution
$\gamma:[0,+\infty)\rightarrow (R^{d})^{N}$ starting from $y_{i}$ at
$t=0$ and asymptotic to $y_{0}$ for $t\rightarrow+\infty$. This
solution is a minimizer of the Lagrangian action with fixed ends in
every compact interval contained in $[0,+\infty)$ and it is
collision-free for $t>0$.}

Referring to the parabolic and hyperbolic orbits, there are some
equivalent definition. These two kind of orbits are both called
hyperbolic-like orbits by Felmer and Tanaka in \cite{3}, which
satisfy
\begin{eqnarray}
|u(t)|\longrightarrow\infty\ \ \ \mbox{as}\ \ \
t\longrightarrow\pm\infty.\label{5}
\end{eqnarray}
The difference between parabolic and hyperbolic orbits is the total
energy $H$ which is shown in $(\mbox{ii})$ and $(\mbox{iii})$.

Motivated by the above papers, we study systems
(\ref{1})$-$(\ref{5}). Under some assumptions, we obtain the
hyperbolic orbits for (\ref{1})$-$(\ref{5}) with $H>0$. Precisely,
we prove the following theorem.

\vspace{0.3cm}{\bf Theorem 1}\ \  {\em Suppose that
$V(x)=\displaystyle\frac{1}{2}\sum_{1\leq i\neq j\leq
N}V_{ij}(x_{i}-x_{j})$ with
$V_{ij}(x_{i}-x_{j})=-\displaystyle\frac{m_{i}m_{j}}{|x_{i}-x_{j}|^{\alpha}}$,
where $\alpha>2$. Then for any $H>0$, there is at least one
hyperbolic orbit for systems (\ref{1})$-$(\ref{5}).}

\section{Variational Settings}

\ \ \ \ \ \ Let us set
\begin{eqnarray*}
&& M=\sum_{i=1}^{N}m_{i},\nonumber\\
&&H^{1}=W^{1,2}(R^{1}/Z,R^{d}),\nonumber\\
&& H^{N}=\{q=(q_{1},\cdots,q_{N})|\ q_{i}\in H^{1},i=1,\cdots,N\}\nonumber\\
&&E_{R}=\{q\in H_N|\ q_{i}(t+1/2)=-q_{i}(t),|q_{i}(0)|=|q_{i}(1)|=R, i=1,\cdots,N\},\nonumber\\
&&\Lambda_{R}=\{q\in E_{R}|\ q_{i}(t)\neq q_{j}(t),\forall t\in
[0,1], \forall i\neq j\}.
\end{eqnarray*}

Here we just use $R$ to denote the Euclidean length of $q_{i}(0)$
and $q_{i}(1)$, $i=1,\cdots,N$. For any $q\in \Lambda_{R}$, it is
easy to check that $\displaystyle\int^{1}_{0}q(t)dt=0$, then by
Poincar$\acute{\mbox{e}}$-Wirtinger's inequality, we obtain the
following equivalent norm in $H^{N}$
\begin{eqnarray}
\|q\|_{H^N}=\left(\int^{1}_{0}\sum_{i=1}^{N}m_{i}|\dot{q}_{i}(t)|^2dt\right)^{1/2}.\nonumber
\end{eqnarray}
Let $L^{\infty}([0,1],(R^{d})^{N})$ be a space of measurable
functions from $[0,1]$ into $(R^{d})^{N}$ and essentially bounded
under the following norm
\begin{eqnarray*}
\ \|q\|_{L^{\infty}([0,1],R^{d}\times\cdots\times
R^{d})}:=\sum_{i=1}^{N}m_{i}\|q_{i}\|_{L^{\infty}([0,1],R^{d})}^{2},
\end{eqnarray*}
where
\begin{eqnarray*}
\|q_{i}\|_{L^{\infty}([0,1],R^{d})}:=esssup\{|q_{i}(t)|:t\in[0,1]\}.
\end{eqnarray*}

Moreover, let $f:\ \Lambda_{R}\rightarrow R^{1}$ be the functional
 defined by
\begin{eqnarray*}
f(q)&=&\frac{1}{2}\int^{1}_{0}\sum_{i=1}^{N}m_{i}|\dot{q}_{i}(t)|^2dt\int^{1}_{0}(H-V(q(t)))dt\\
&=& \mbox{}\frac{1}{2}\|q\|^{2}\int^{1}_{0}(H-V(q(t)))dt
\end{eqnarray*}
Then one can easily check that $f\in C^{1}(\Lambda_{R},R^{1})$ and
\begin{eqnarray*}
\ ( f'(q),q)&=&\|q\|^2\int^{1}_{0}\left(H-V(q(t))-\frac{1}{2}(
\nabla V(q(t)),q(t))\right)dt.
\end{eqnarray*}

Our way to get the hyperbolic orbit is by approaching it with a
sequence of periodic solutions. Firstly, we prove the existence of
the approximate solutions, then we study the limit procedure.

\section{Existence of Periodic Solutions}

\ \ \ \ \ \ The approximate solutions are obtained by the
variational minimization methods. We need the following lemma which
is proved by A. Ambrosetti and V. C. Zelati in \cite{1}.

\vspace{0.3cm}{\bf Lemma 3.1(See\cite{1})}\ {\em Let
$f(q)=\displaystyle\frac{1}{2}\int^{1}_{0}\sum_{i=1}^{N}m_{i}|\dot{q}_{i}(t)|^2dt\int^{1}_{0}(H-V(q(t)))dt$
and $\tilde{q}\in H^{N}$ be such that $f^{'}(\tilde{q})=0$,
$f(\tilde{q})>0$. Set
\begin{eqnarray*}
T^{2}=\frac{\displaystyle\frac{1}{2}\displaystyle\int^{1}_{0}\sum_{i=1}^{N}m_{i}|\dot{\tilde{q}}_{i}(t)|^{2}dt}{\displaystyle\int^{1}_{0}(H-V(\tilde{q}(t))dt}.
\end{eqnarray*}
Then $\tilde{u}(t)=\tilde{q}(t/T)$ is a non-constant $T$-periodic
solution for (\ref{1}) and (\ref{2}).}

\vspace{0.3cm}{\bf Lemma 3.2(Palais\cite{13})}\ {\em Let $\sigma$ be
an orthogonal representation of a finite or compact group $G$ in the
real Hilbert space $H$ such that for any $\sigma\in G$,
\begin{eqnarray*}
f(\sigma\cdot x)=f(x),
\end{eqnarray*}
where $f\in C^{1}(H,R^{1})$. Let $S=\{x\in H|\sigma x=x,
\forall\sigma\in G\}$, then the critical point of $f$ in $S$ is also
a critical point of $f$ in $H$.}

Lemma 3.2 guarantee that the critical points of $f$ in $\Lambda_{R}$
are still the critical points in the whole space.

\vspace{0.3cm}{\bf Lemma 3.3(Translation Property\cite{8})}\ {\em
Suppose that, in domain $D\subset R^{d}$, we have a solution
$\phi(t)$ for the following differential equation
\begin{eqnarray*}
x^{(n)}+F(x^{(n-1)},\cdots,x)=0,
\end{eqnarray*}
where $x^{(k)}=d^{k}x/dt^{k}$, $k=0,1,\cdots,n$, $x^{(0)}=x$. Then
$\phi(t-t_{0})$ with $t_{0}$ being a constant is also a solution.}

\vspace{0.3cm}{\bf Lemma 3.4}\ {\em Let $E$ be a Banach space,
$f\not\equiv+\infty :E\rightarrow R^{1}$ a functional bounded from
below and $c=\inf_{E}f$. If f satisfies the $(CPS)_{c}$ condition
and
\begin{eqnarray*}
f(x_{j})\rightarrow +\infty,\ \ \ \mbox{as}\ \ x_{j}\rightharpoonup
x_{0}\in \partial\Lambda_{R}.
\end{eqnarray*}
then f attains its infimum on $E$.}

The proof of Lemma 3.4 can easily be obtained from Ambrosetti and
Zelati in \cite{20}. In order to prove that the minimizing sequence
converges on $\Lambda_{R}$, we need to introduce the $Gordon's\
Strong\ Force$ condition.

\vspace{0.3cm}{\bf Definition 3.5(Gordon\cite{19})}\ \ {\em $V$ is
said to satisfies the $Gordon's\ Strong\ Force$ condition, if there
exists a neighborhood $\mathcal{N}$ of 0 and a function $U\in
C^{1}(R^{d}\setminus\{0\},R^{1})$ such that}

$(\mbox{i})$\ $\lim_{x\to 0}U(x)=-\infty$;

$(\mbox{ii})$\ $-V(x)\geq |U'(x)|^{2}$ for every $x\in \mathcal{N}$\
$\setminus\{0\}$,

{\em with}
\begin{eqnarray*}
\int^{1}_{0}V(x_{j})dt\rightarrow-\infty,\ \ \ \forall\
x_{j}\rightharpoonup x\in \partial \Lambda_{R}.
\end{eqnarray*}

\vspace{0.3cm}{\bf Lemma 3.6}\ \ Suppose $V_{ij}$ satisfies  the
condition in Theorem 1, then $V_{ij}$ satisfies the $Gordon's\
Strong\ Force$ condition.

\vspace{0.3cm}{\bf Proof.}\ Let $\phi(r)=-V_{ij}(re)r^{2}$, where
$r=|x|$, $e=x/|x|$, then we have
\begin{eqnarray*}
\phi'(r)=-r(2V_{ij}(re)+(\nabla V_{ij}(re),re)).
\end{eqnarray*}
It follows from the definition of $V_{ij}$ that, there exists a
constant $\delta>0$ such that
\begin{eqnarray*}
\phi'(r)\leq0\ \ \ \mbox{for all}\ \ \ 0<r\leq\delta.
\end{eqnarray*}
Since $V_{ij}\in C^{1}(R^{D}\setminus\{0\},R^{1})$, we get
\begin{eqnarray*}
\phi(r)&\geq&\phi(\delta)=-V_{ij}(\delta e)\delta^{2}\geq
\delta^{2}\min_{|x|=\delta}(-V_{ij}(x)).
\end{eqnarray*}
It follows from the definition of $\phi$ that there exists a
constant $C>0$ such that
\begin{eqnarray*}
-V_{ij}(x)\geq\frac{C}{|x|^{2}}\ \ \ \mbox{for all}\ \ \
0<r\leq\delta.
\end{eqnarray*}
We set $U(x)=\sqrt{C}\ln|x|$, then by some calculation, we obtain
\begin{eqnarray*}
\lim_{x\to 0}U(x)=-\infty\ \ \ \mbox{and}\ \ \
-V_{ij}(x)\geq|U'(x)|^{2}\ \ \ \mbox{for all}\ \ \ 0<r\leq\delta,
\end{eqnarray*}
which proves this lemma.

\vspace{0.3cm}{\bf Lemma 3.7}\ {\em Suppose the conditions of
Theorem 1 hold, then for any $R>0$, there exists at least one
periodic solution in $\Lambda_{R}$ for the following systems
\begin{equation}
   m_{i}\ddot{u}_{i}(t)+\nabla_{u_{i}} V(u_{1}(t),\cdots,u_{N}(t))=0 \ \ (1\leq i\leq N), \ \ \ \
   \forall\ t\in\left(-\frac{T_{R}}{2},\frac{T_{R}}{2}\right)\label{16}
\end{equation}
with
\begin{equation}
   \frac{1}{2}\sum_{i=1}^{N} m_{i}|\dot{u}_{i}(t)|^{2}+ V(u_{1}(t),\cdots,u_{N}(t))=H,\ \ \ \
   \forall\ t\in\left(-\frac{T_{R}}{2},\frac{T_{R}}{2}\right)\label{17}.
\end{equation}}

\vspace{0.3cm}{\bf Proof.}\ We notice that $H^{N}$ is a reflexive
Banach space and $E_{R}$ is a weakly closed subset of $H^{N}$. Since
total energy $H>0$, we obtain that
\begin{eqnarray}
f(q)&=&\frac{1}{2}\|q\|^{2}\int^{1}_{0}(H-V(q(t))dt\geq\frac{H}{2}\|q\|^{2},\label{9}
\end{eqnarray}

Then, we conclude that for every $R>0$ there exists a minimizer
$q_{R}\in E_{R}$ such that
\begin{eqnarray}
f'(q_R)=0,\ \ \ \ f(q_R)=\inf_{q\in E_{R}} f(q)>0.\label{18}
\end{eqnarray}
Furthermore, we need to prove that $q_{R}\in\Lambda_{R}$ which means
$q_{R}$ has no collision for any $R>0$. Suppose that
$\{q_{j}\}_{j\in N}$ is the minimizing sequence, then if $q_{R}$ has
collision, which means $q_{R}\in
\partial\Lambda_{R}=\{q_{R}\in E_{R}|\ \exists\ t'\in [0,1]\ st.\
q_{R}(t')=0\}$, we can prove that
\begin{eqnarray}
f(q_{j})\rightarrow +\infty,\ \ \ \mbox{as}\ \
j\rightarrow+\infty.\label{10}
\end{eqnarray}
To prove this fact, there are two cases needed to be discussed.

{\bf Case 1.}\ If $q_R=$constant, it follows from $q_{R}\in
\partial\Lambda_{R}$ that $q_{R}\equiv0$, which is a contradiction,
since $|q_{R}(0)|=|q_{R}(1)|=R$.

{\bf Case 2.}\ If $q_{R}\neq$constant, we have
$\|q_{R}\|^{2}=\displaystyle\int^{1}_{0}\sum_{i=1}^{N}m_{i}|\dot{q}_{R,i}(t)|^2dt>0$,
otherwise by $q_{R,i}(t+1/2)=-q_{R,i}(t)$, we can deduce
$q_{R,i}\equiv0$ which is a contradiction. Then by the
weakly-lower-semi-continuity of norm, we have
\begin{eqnarray*}
\liminf_{j\rightarrow\infty}\|q_{j}\|\geq \|q_{R}\|>0.
\end{eqnarray*}
Then by Lemma 3.4, (\ref{10}) holds.

Let $Q=(Q_{1},Q_{2},\cdots,Q_{N})$ with $Q_{i}(t)=R[\xi_{i}\cos2\pi
(t+\frac{i}{N})+\eta_{i}\sin2\pi (t+\frac{i}{N})]\in \Lambda_R$,
where $\xi_{i}, \eta_{i} \in R^{d}\setminus\{0\}$,
$|\xi_{i}|=|\eta_{i}|=1$, $(\xi_{i},\eta_{i})=0$,  which implies
that $|Q(t)|=RN$, $\|Q\|^{2}=4\pi^{2} R^{2}M$, hence
\begin{eqnarray*}
f(Q)&=&2\pi^{2}R^{2}M\left(H-\int^{1}_{0}V(Q(t))dt\right)
\end{eqnarray*}
Since $V_{ij}\in C^{1}(R^{D}\setminus\{0\},R^{1})$, then there
exists a constant $M_{1,R}> 0$ such that $|V(Q(t))|\leq M_{1,R}$. We
obtain that
\begin{eqnarray}
f(q_R)\leq f(Q)\leq M_{2,R}\label{22}
\end{eqnarray}
for some $M_{2,R}>0$, but (\ref{22}) contradicts with (\ref{10}) for
any fixed $R>0$. Then we can see that $q_{R}\in \Lambda_{R}$ has no
collision.

By Lemma 3.4, we conclude that for every $R>0$ there exists
$q_{R}\in\Lambda_{R}$ such that
\begin{eqnarray}
f'(q_{R})=0,\ \ \ \ f(q_{R})=\inf_{q\in \Lambda_{R}}
f(q)>0.\label{18}
\end{eqnarray}

Let
\begin{eqnarray}
T_{R}^{2}=\frac{\displaystyle\frac{1}{2}\displaystyle\int^{1}_{0}\sum_{i=1}^{N}m_{i}|\dot{q}_{R,i}(t)|^{2}dt}{\displaystyle\int^{1}_{0}(H-V(q_{R}(t))dt}.\label{21}
\end{eqnarray}

Then by Lemma 3.1$-$ Lemma 3.4, we obtain that
$u_{R}(t)=q_{R}(\frac{t+\frac{T_{R}}{2}}{T_{R}}):\left(-\frac{T_{R}}{2},\frac{T_{R}}{2}\right)\rightarrow
\Lambda_{R}$ is a $T_{R}$-periodic solution for systems (\ref{16})
and (\ref{17}). The Lemma is proved, which is
\begin{equation}
   m_{i}\ddot{u}_{R,i}(t)+\nabla_{u_{i}} V(u_{R}(t))=0\label{32},
\end{equation}
with
\begin{equation}
   \frac{1}{2T_{R}^{2}}\sum_{i=1}^{N}m_{i}|\dot{q}_{R,i}(t)|^{2}+ V(q_{R}(t))=H\label{33}.
\end{equation}

We have proved the existence of periodic solutions for systems
(\ref{16})$-$(\ref{17}) for any $R>0$, in order to get the
hyperbolic solutions, we need to let $R\rightarrow+\infty$ which
need blowing-up arguments.

\section{Blowing-up Arguments}

\ \ \ \ \ \ Subsequently, we need to show that the distance between
any two bodies can not diverge to infinity uniformly as
$R\rightarrow+\infty$. Moreover, we prove the following lemma.

\vspace{0.3cm}{\bf Lemma 4.1}\ {\em Suppose that
$u_{R}(t):\left[-\frac{T_{R}}{2},\frac{T_{R}}{2}\right]\rightarrow
\Lambda_{R}$ is the solution obtained in Lemma 3.7, then $u_{R}$ has
no collisions. Moreover, we obtain that there exist constants
$C_{0}$, $C_{1}>0$ independent of $R$ such that
\begin{eqnarray*}
C_{1}\geq\min_{i\neq j,
t\in\left[-\frac{T_{R}}{2},\frac{T_{R}}{2}\right]}|u_{R,i}(t)-u_{R,j}(t)|\geq
C_{0}\ \ \ \mbox{for all}\ \ \ R>0.
\end{eqnarray*} }

\vspace{0.3cm}{\bf Proof.}\ We set
\begin{eqnarray*}
\varphi_{R}(u_{R})=\min_{i\neq j,
t\in\left[-\frac{T_{R}}{2},\frac{T_{R}}{2}\right]}|u_{R,i}(t)-u_{R,j}(t)|^{\alpha}.
\end{eqnarray*}
By $f'(q_{R})=0$ and $\|q_{R}\|\neq0$, we obtain that
\begin{eqnarray*}
\int^{\frac{T_{R}}{2}}_{-\frac{T_{R}}{2}}2(H-V(u_{R}(t)))-(\nabla
V(u_{R}(t)),u_{R}(t))dt=0.
\end{eqnarray*}
Then we can deduce that there exists
$t_{0}\in\left[-\frac{T_{R}}{2},\frac{T_{R}}{2}\right]$ such that
\begin{eqnarray*}
2(H-V(u_{R}(t_{0})))-(\nabla V(u_{R}(t_{0})),u_{R}(t_{0}))\leq0,
\end{eqnarray*}
which implies that
\begin{eqnarray*}
2H&\leq&2V(u_{R}(t_{0}))+(\nabla V(u_{R}(t_{0})),u_{R}(t_{0}))\nonumber\\
&\leq& \mbox{} -(\alpha-2)V(u_{R}(t_{0}))\nonumber\\
&\leq& \mbox{}\frac{(\alpha-2)}{2}\sum_{1\leq i\neq j\leq
N}\frac{m_{i}m_{j}}{|u_{R,i}(t_{0})-u_{R,j}(t_{0})|^{\alpha}}\nonumber\\
&\leq& \mbox{} \frac{(\alpha-2)\sum_{1\leq i\neq j\leq
N}m_{i}m_{j}}{2\varphi_{R}(u_{R})}.
\end{eqnarray*}
Since $H>0$, there exists a constant $M_{7}$ such that
\begin{eqnarray*}
\varphi_{R}(u_{R})\leq M_{7}>0.
\end{eqnarray*}

On the other hand, we can deduce that
\begin{eqnarray}
0&=&\int^{1}_{0}2(H-V(q_{R}(t)))-(\nabla
V(q_{R}(t)),q_{R}(t))dt\nonumber\\
&=& \mbox{}
\int^{1}_{0}2H+(\alpha-2)V(q_{R}(t))dt\nonumber\\
&=& \mbox{} \int^{1}_{0}2H-(\alpha-2)\sum_{1\leq i< j\leq
N}\frac{m_{i}m_{j}}{|q_{R,i}(t)-q_{R,j}(t)|^{\alpha}}dt. \label{36}
\end{eqnarray}
Set
\begin{eqnarray*}
J_{R}=\{t\in[0,1]|\exists\ i_{0}\neq j_{0}\ st.\
|q_{R,i_{0}}(t)-q_{R,j_{0}}(t)|\rightarrow0\ \mbox{as}\
R\rightarrow+\infty \}.
\end{eqnarray*}

As Saari and Hulkower stated in \cite{25}, if $J_{R}\neq\emptyset$,
i.e. $\exists\ t_{0}\in J_{R}$ for some $i_{0}\neq j_{0}$, we have
the following asymptotic estimates, for some $A>0$
\begin{eqnarray}
|q_{R,i_{0}}(t)-q_{R,j_{0}}(t)|^{-\alpha}\sim
A|t-t_{0}|^{-\frac{2\alpha}{\alpha+2}}+o(|t-t_{0}|^{-\frac{2\alpha}{\alpha+2}})\
\ \ \mbox{as}\ \ t\rightarrow t_{0}.\label{38}
\end{eqnarray}
Set a sequence $\{t_{n}\}\subset (0,1)$ such that $t_{n}\rightarrow
0$ as $n\rightarrow+\infty$. By (\ref{38}), there exists a $B>0$
such that
\begin{eqnarray*}
|q_{R,i_{0}}(t)-q_{R,j_{0}}(t)|^{-\alpha}\geq
\frac{1}{2}A|t-t_{0}|^{-\frac{2\alpha}{\alpha+2}}-1\ \ \ \mbox{for
all}\ \ t\in[t_{0}+t_{n},t_{0}-t_{n}],\ n>B,
\end{eqnarray*}
which implies that
\begin{eqnarray}
 \frac{2H}{\alpha-2}
&\geq& \int^{1}_{0}\sum_{1\leq i< j\leq
N}\frac{m_{i}m_{j}}{|q_{R,i}(t)-q_{R,j}(t)|^{\alpha}}dt\nonumber\\
&\geq& \mbox{}\int^{t_{0}+t_{n}}_{t_{0}-t_{n}}\frac{m_{i_{0}}m_{j_{0}}}{|q_{R,i_{0}}(t)-q_{R,j_{0}}(t)|^{\alpha}}dt\nonumber\\
&\geq&
\mbox{}\frac{1}{2}Am_{i_{0}}m_{j_{0}}\int^{t_{0}+t_{n}}_{t_{0}-t_{n}}|t-t_{0}|^{-\frac{2\alpha}{\alpha+2}}dt-2m_{i_{0}}m_{j_{0}}t_{n}\nonumber\\
&=&
\mbox{}\frac{1}{2}Am_{i_{0}}m_{j_{0}}\int^{t_{n}}_{-t_{n}}|s|^{-\frac{2\alpha}{\alpha+2}}ds-2m_{i_{0}}m_{j_{0}}t_{n}.\label{39}
\end{eqnarray}
Since $\alpha>2$, we deduce that
$\displaystyle\frac{2\alpha}{\alpha+2}>1$, which means that
$\displaystyle\int^{t_{n}}_{-t_{n}}|s|^{-\frac{2\alpha}{\alpha+2}}ds=+\infty$.
This contradicts (\ref{39}). Then $q_{R}$ has non collision
uniformly as $R\rightarrow+\infty$. The same with $u_{R}$.

\section{Existence of Hyperbolic Solutions}

\ \ \ \ \ \ We set two constants $1-1/d_{2}>d_{1}-1>0$ such that
\begin{eqnarray*}
d_{1}R>R>\frac{R}{d_{2}}>0
\end{eqnarray*}
and
\begin{eqnarray*}
S=\left\{t\in\left[-\frac{T_{R}}{2}, \frac{T_{R}}{2}\right]|\
|u_{R,i}(t)|=d_{1}R\ \ \mbox{or}\ \ |u_{R,i}(t)|=\frac{R}{d_{2}},\ \
\mbox{for any}\ \ i=1,\cdots,N\right\}\neq\emptyset.
\end{eqnarray*}

\vspace{0.3cm}{\bf Lemma 5.1}\ {\em Suppose that $u_{R}(t)$ is the
solution for (\ref{16}) and (\ref{17}) obtained in Lemma 3.7 and set
$\displaystyle-\frac{T_{R}}{2}<t_{-}\leq
t_{+}<\displaystyle\frac{T_{R}}{2}$ such that
\begin{eqnarray*}
t_{+}=\sup S\ \ \ \mbox{and}\ \ \ t_{-}=\inf S.
\end{eqnarray*}

 Then we have that}
\begin{eqnarray*}
\frac{T_{R}}{2}-t_{+}\rightarrow+\infty,\ \ \
t_{-}+\frac{T_{R}}{2}\rightarrow+\infty\ \ \ \mbox{as}\ \
R\rightarrow+\infty.
\end{eqnarray*}

\vspace{0.3cm}{\bf Proof.}\ By the definition of $u_{R}(t)$ we have
that
\begin{eqnarray*}
\left|u_{R,i}\left(-\frac{T_{R}}{2}\right)\right|=\left|u_{R,i}\left(\frac{T_{R}}{2}\right)\right|=R\
\ \ \mbox{for any}\ \ \ i=1,\cdots,N.
\end{eqnarray*}
From the definition of $V$, and the definitions of $t_{+}$, $d_{1}$
and $d_{2}$, we have
\begin{eqnarray}
\int^{\frac{T_{R}}{2}}_{t_{+}}\sqrt{H-V(u_{R}(t))}\sum_{i}\sqrt{m_{i}}|\dot{u}_{R,i}(t)|dt&\geq&\sqrt{H}\sum_{i}\sqrt{m_{i}}\left|\int^{\frac{T_{R}}{2}}_{t_{+}}\dot{u}_{R,i}(t)dt\right|\nonumber\\
&=& \mbox{}
\sqrt{H}\sum_{i}\sqrt{m_{i}}\left|u_{R,i}\left(\frac{T_{R}}{2}\right)-u_{R,i}(t_{+})\right|\nonumber\\
&\geq& \mbox{}
\sqrt{H}\sum_{i}\sqrt{m_{i}}\left(d_{1}-1\right)R.\label{31}
\end{eqnarray}
Then by Lemma 4.1, we can deduce that there exists a constant
$M_{1}>0$ independent of $R$ such that
\begin{eqnarray}
|V(u_{R}(t))|\leq M_{1}\ \ \ \mbox{for}\ \ \mbox{all}\ \
t\in\left[-\frac{T_{R}}{2},\frac{T_{R}}{2}\right].\label{24}
\end{eqnarray}
It follows from the definition of $t_{+}$ and (\ref{17}) that
\begin{eqnarray*}
\int^{\frac{T_{R}}{2}}_{t_{+}}\sqrt{H-V(u_{R}(t))}\sum_{i}\sqrt{m_{i}}|\dot{u}_{R,i}(t)|dt&\leq&\int^{\frac{T_{R}}{2}}_{t_{+}}\sqrt{N}\sqrt{H-V(u_{R}(t))}\sqrt{\sum_{i}m_{i}|\dot{u}_{R,i}(t)|^{2}}dt\nonumber\\
&=& \mbox{}
\int^{\frac{T_{R}}{2}}_{t_{+}}\sqrt{2N}(H-V(u_{R}(t)))dt\nonumber\\
&\leq& \mbox{} \sqrt{2N}(H+M_{1})\left(\frac{T_{R}}{2}-t_{+}\right).
\end{eqnarray*}
Combining (\ref{31}) with the above estimate, we obtain that
\begin{eqnarray*}
\frac{T_{R}}{2}-t_{+}\rightarrow+\infty,\ \ \ \mbox{as}\ \
R\rightarrow+\infty.
\end{eqnarray*}
The limit for $t_{-}+\frac{T_{R}}{2}$ can be obtained in the similar
way. The proof is completed.

We can fix $t^{*}$ such that $t_{+}\geq t^{*}\geq t_{-}$, which
implies that
\begin{eqnarray*}
-\frac{T_{R}}{2}+t^{*}\rightarrow-\infty,\ \ \mbox{}\ \
\frac{T_{R}}{2}+t^{*}\rightarrow+\infty\ \ \mbox{as}\ \
R\rightarrow\infty.
\end{eqnarray*}

Then we set
\begin{eqnarray*}
u_{R}^{*}(t)=u_{R}(t-t^{*}).
\end{eqnarray*}
Since $u^{*}_{R}$ is a solution for systems
\begin{equation}
   m_{i}\ddot{u}^{*}_{i}(t)+\nabla_{u_{i}} V(u^{*}_{1}(t),\cdots,u^{*}_{N}(t))=0 \ \ (1\leq i\leq
   N),\label{34}
\end{equation}
with
\begin{equation}
   \frac{1}{2}\sum_{i=1}^{N} m_{i}|\dot{u}^{*}_{i}(t)|^{2}+
   V(u^{*}_{1}(t),\cdots,u^{*}_{N}(t))=H\label{35}
\end{equation}
for all
$t\in\left(-\frac{T_{R}}{2}+t^{*},\frac{T_{R}}{2}+t^{*}\right)$. It
follows from (\ref{24}) and (\ref{35}) that there is a constant
$M_{2}$ independent of $R$ such that
\begin{eqnarray*}
 |\dot{u}^{*}_{R}(t)|\leq M_{2}\ \ \ \mbox{for all}\ \
 t\in\left(-\frac{T_{R}}{2}+t^{*},\frac{T_{R}}{2}+t^{*}\right).
\end{eqnarray*}
which implies that
\begin{eqnarray*}
\ |u_{R}^{*}(t_{1})-u_{R}^{*}(t_{2})| \leq
\left|\int^{t_{1}}_{t_{2}}\dot{u}_{R}^{*}(s)ds\right| \leq \mbox{}
\int^{t_{1}}_{t_{2}}\left|\dot{u}_{R}^{*}(s)\right|ds \leq
M_{2}|t_{1}-t_{2}|
\end{eqnarray*}
for each $R>0$ and $t_{1}, t_{2} \in
\left(-\frac{T_{R}}{2}+t^{*},\frac{T_{R}}{2}+t^{*}\right)$, which
shows $\{u_{R}^{*}\}$ is equicontinuous. Then there is a subsequence
$\{u_{R}^{*}\}_{R>0}$ converging to $u_{\infty}$ in
$C_{loc}(R^{1},(R^{d})^{N})$. Then there exists a function
$u_{\infty}(t)$ such that
\begin{eqnarray*}
&&(\mbox{i})\ u_{R}^{*}(t)\rightarrow u_{\infty}(t)\ \mbox{in}\ C_{loc}(R^{1},(R^{d})^{N})\nonumber\\
&&(\mbox{ii})|u_{\infty}(t)|\rightarrow +\infty\ \mbox{as}\
|t|\rightarrow+\infty
\end{eqnarray*}
and $u_{\infty}(t)$ satisfies systems $(1)-(2)$.

From the above lemmas, we have proved there is at least one
hyperbolic solution for $(1)-(2)$ with $H>0$. We finish the proof.
$\Box$

\end{document}